# Test2Vec: An Execution Trace Embedding for Test Case Prioritization

EMAD JABBAR, University of Calgary, Canada

SOHEILA ZANGENEH, University of Calgary, Canada

HADI HEMMATI, University of Calgary, Canada

ROBERT FELDT, Chalmers University of Technology, Sweden

Most automated software testing tasks, such as test case generation, selection, and prioritization, can benefit from an abstract representation of test cases. Traditional test case representation is typically an encoding of test cases based on their code coverage. For instance, which statements a test case covers can be seen as features, representing the test case. Specification-level criteria (such as statement or path coverage in a state machine) can replace (or complement) code coverage to better represent test cases with respect to their behaviour, since such abstractions are typically closer to the software requirements. However, specification-based testing, which encodes test cases using specification-level representations is often not cost-effective. Thus, most current automated testing techniques rely on code-based representations of test cases.

In this paper, we hypothesize that execution traces of the test cases, as an abstraction of their behaviour, can be leveraged to better represent test cases compared to static code-based methods. To validate this hypothesis, we propose a novel embedding approach, Test2Vec, that maps test execution traces, i.e. sequences of method calls with their inputs and return values, to fixed-length, numerical vectors. We evaluate this representation in test case prioritization (TP), as a common and important automated testing task. Our default TP method is based on similarity of the embedded vectors to vectors of failing test cases from previous system revisions. We also study an alternative TP approach based on the diversity of test vectors. Finally, we propose a simple prediction method to decide which TP to choose, for a given test suite, based on its vectors and the historical test cases. We compare Test2Vec to traditional coverage-based embeddings, as non-trace-embedding baseline. We also compare it with alternative trace-embedding baselines, including simple one-hot encoding, state-of-the-art neural program embedding (CodeBert), and a related work on trace embedding (based on the LSTM model), all in the context of a TP task. The empirical study is based on 250 real faults and 703, 353 seeded faults (mutants) over 250 revisions of 10 open-source Java projects from Defects4J, with a total of over 1, 407, 206 execution traces. Results show that our best Test2Vec-based TP improves on the best alternative (CodeBERT), by 41.80% in terms of the median normalized rank of the first failing test case metric. Furthermore, our default Test2Vec-based TP (Test2Vec classifier-TP) outperforms traditional code coverage based approaches by 25.05% and 68.39% in terms of median APFD and median normalized rank of the first failing test case.

## 1 INTRODUCTION

Code coverage is one of the most common test adequacy metrics and can ensure that different source code elements, such as statements and conditions, are executed at least once during test execution [1, 50, 83]. It is one of the main targets to optimize when designing an automated test generation [29] or prioritization [38] technique. However, the current code-based coverage metrics are weak criteria in evaluating a test case/suite in terms of detecting bugs. A fully covered source code is not necessarily bug free; specification-related bugs tend to be missed by code coverage metrics [37, 46]. Other alternatives such as fuzzing [32, 84] and input data or test case diversification [13, 23] are also limited in covering software behaviour thoroughly, since they look at the software under test more like a black-box and do not leverage the underlying source code, execution traces, or specification artifacts. Specification-level criteria, which are defined on the specified requirements (e.g., modeled as a state machine), can help fill the gap. For example, using model-based testing [40, 41] one can represent desired behaviour of the system under test and generate and prioritize tests accordingly. However, these techniques are often costly, since they incur the overhead of building models, continuously maintaining them, and the skill set to build, update, and use the models [2].

To tackle the overhead of model-driven software engineering, in general, there has been a lot of progress in reverse engineering different types of models [49, 53]. Creating a behavioral model of a system under test can help identify what parts of the system behavior has been covered by the existing tests and what parts need more testing. However, most of these techniques are not designed with a testing use case in mind. For example, a large body of the specification mining work only abstract a finite state machine (FSM) from the sequence of method call invocations, which is limited since, in testing, input values are often more critical and their diversity and coverage can improve testing [24, 26, 42].

The goal of this paper is to propose a novel methodology to represent the run-time behavior of test cases that is optimized for testing, and that requires little to no additional manual modeling before use. In other words, a test case representation that can make distinctions between fail and pass behavior (i.e., the ultimate goal of automated testing tasks such as test generation, selection, and prioritization). This requires a rich representation of test cases. For instance, in this paper, we emphasize that input and output variables are important in testing and the behaviour of the test case should not be represented only using a sequence of method calls [26, 63, 64]. However, including test inputs and outputs increases the dimensionality of the representation space. Therefore a simple encoding such as the common one-hot-encoding approach is not applicable, and a more advanced model is needed to encode information-rich test execution traces.

Recently, there has been a lot of progress in applying sequence learning (particularly embedding models) in software engineering. Most of these works apply an embedding on source code (code embedding) to capture semantic information about the code. However, their focus is mostly on syntactic, and thus static, aspects of the code, while from the testing perspective the dynamic behavior of the program is of main interest. Recent works have shown that embedding dynamic execution traces of programs is helpful for tasks such as program repair [79, 80] and test oracle generation [75, 76].

Therefore, in this study, we propose a neural sequence learning model called "Test2Vec", which can create general representations of test case behavior using historical snapshots of a project. This can potentially be used for multiple downstream testing tasks, similar to how such embedding methods in natural language processing (NLP) are successfully used in several NLP tasks [52, 82]. Our model uses a well-known transformer-based model, called CodeBERT, to learn vectorized representations of sequences of method calls, outputs, and input parameters, separately. Finally, the model also consists of a BiLSTM layer [31] to act as a classifier and learn the testing-specific characteristics of the sequences, e.g. as pass or fail, based on their similarity to the historical failures. BiLSTM layers are chosen over a simpler classifier layer such as MLP, since they can better learn from the before and after context in an embedding sequence.

To evaluate our approach, we conduct a large experiment on 250 faulty releases from 10 open-source Java projects from the Defect4J [48] benchmark dataset, with a total of $758,396$ execution traces. We use test case prioritization (TP) as a common downstream task to show the practicality and real-world usage of Test2Vec. Our Test2Vec-based TP leverages two heuristics for ranking test cases: (a) the similarity to previously failing test cases, which is motivated from regression testing literature [62] and implemented as a classifier, and (b) the dissimilarity to other test cases in the test suite, which is motivated by test case diversification literature [24, 59] and implemented as an anomaly detection method.

As baselines in our experimental evaluation, we use five different TP techniques from four different categories: (a) two traditional coverage-based TPs as simple approaches representing state-of-practice, (b) a basic trace encoding technique (One-Hot encoding),(c) an alternative state-of-the-art embedding approach (CodeBERT) as the best embedding already applied on other domains in software engineering, and (c) an LSTM-based approach from the literature [77], as the state-of-art method for classifying test case execution as pass or fail.

We explore two main research questions (RQs) in the experiments. In RQ1, we focus on showing the overall effectiveness of our embedding architecture compared to alternatives, both as simple coverage-based baselines and most advanced embedding architectures. In RQ2, we look at our down-stream task more carefully and explore two potential solutions for the prioritization task that can be implemented by the Test2Vec embedded vectors: classification and anomaly detection. Our research questions and their sub-RQs are as follows:

- RQ1) How effective is Test2Vec embedding for test prioritization, when the tests are ranked using a classifier trained on historical data?
  - RQ1-1) How effective is Test2Vec classifier-TP compared to code coverage-based prioritization (state of the practice)?
  - RQ1-2) How effective is Test2Vec classifier-TP compared to a basic and a state-of-art code and execution trace embedding technique from literature?
- RQ2) Which of the two prioritization heuristics ("similarity to past failing tests" OR "test diversification") is better to be used with Test2Vec, for a test case prioritization task?
  - RQ2-1) How effective is test diversification using Test2Vec embedding (Test2Vec diversification-TP), for test prioritization?
  - RQ2-2) Can combining diversification and history-based classification (Test2Vec combined-TP) outperform them individually?

To evaluate RQ1, we use both the median and mean normalized rank of the first failing test (FFR) and the well-known APFD measures. The FFR metrics are measured based on real faults collected from Defect4J dataset. APFD however is calculated based on mutation analysis. Overall, RQ1 results show that our default TP technique (Test2Vec classifier-TP) leads to significant improvements, in terms of FFR and APFD. The average (and median) FFR relative improvements over baselines in RQ1 are: compared to line coverage 68.89% (66.15%), to branch coverage 70.28% (73.43%), to one-hot encoding model 54.16% (53.24%), to CodeBERT 54.16% (53.24%), and compared to the LSTM-based model 34.54% (40.70%).

The APFD average (median) relative improvements over baselines, RQ1 results, are: compared to line coverage 26.42% (29.51%), compared to branch coverage 28.72% (30.19%), compared to one-hot model 15.92% (18.75%), compared to CodeBERT 3.81% (3.57%), and compared to the LSTM-based model 18.12% (20.00%).

RQ2 is only assessed based on real-faults (using FFR), since diversity-based approaches are not effective when bugs are seeded artificially (i.e., they don't look like anomalies anymore – we will explain this more in the design section). Our results show that in 23.81% of the times a diversification approach leads to better results than a classifier-based approach. Finally, we show that the ensemble method can improve the results of Test2Vec classifier-TP (from RQ1) by 9.5% and 11.2%, for median and mean of FFR and in 47.61% of the versions studied Test2Vec combined-TP was better or the same as Test2Vec classifier-TP.

The contributions of this paper can be summarized as:

- Proposing a novel embedding approach (Test2Vec) specialized for test case behaviour representation that considers the sequence of method calls alongside with their inputs and outputs.
- Proposing a test case prioritization method, based on the behavioural similarity of current test cases to historical failures' behaviour, leveraging Test2Vec embedding.

- Proposing a diversity-based test case prioritization method that works as an anomaly detection on the behaviour space of a test suite (set of vectors representing test cases using Test2Vec).
- Proposing a novel test case prioritization method that combines both of the above heuristics.
- Conducting a large-scale experiment, comparing the proposed prioritization methods with both traditional coverage-based approaches, as well assessing the Test2Vec embedding architecture compared to the state-of-the-art code and execution trace embedding models, int he context of test cases prioritization down-stream task.

All the source code and datasets of this study are available in a public repository*, for replication.

## 2 MOTIVATION

The hypotheses of this study is that an execution trace-based embedding can better represent test cases compared to traditional code-driven representations (e.g., code coverage). Essentially, we argue that a good summary of the **dynamic** behavior of the program during testing, as captured by the traces, has rich information and can thus better represent the test cases in later tasks. To quantify "better" on a concrete task, in the evaluation section we use TP as a measurable real-world problem. In this section, however, we motivate our approach with a simple and intuitive example. This example is based on an artificial code snippet and its test cases.

Listing 1 shows an example function called *formula* that computes $\frac{b^e}{c}$. One of the actual requirements for its sub-function named *power* is that $e >= 0$, otherwise the function should return -1. But this requirement has been missed in analysis, the code does not implement it, and, thus, the code is faulty. The *check* function is in charge of verifying that all entry requirements are fulfilled. But it contains a bug and does not check whether $e$ is greater or equal to 0.

Assume there are two test cases written for the *formula* function. These two tests have the same branch coverage and statement coverage, while only *test2* fails and triggers the bug. The execution traces (excluding the assertions) for these two tests are as follows:

$$Trace1 : formula(2, 0, 3), power(2, 0), check(2, 0)$$

$$Trace2 : formula(2, -1, 3), power(2, -1), check(2, -1)$$

These tests have the same sequence of method calls with different inputs and only one of them fails.

What we observe from this example is first the code coverage metrics (at least the branch and statement coverage used here) are not enough to make distinction between a failing and a passing test case behavior. Also, we see that the typical behavior representation in the testing domain, which is a sequence of method calls (that is what most related work in testing and specification mining use), will also, like coverage-based representations, be ineffective (zero distinction between pass and fail). Therefore, a better representation is called for, for instance by considering both the sequence of method calls and the input values. Potentially, additional information can also be utilized such as output values.

Motivated by related work such as code embedding approaches, which are, however, limited to source code and not execution traces [6], and neural fuzz testing, which is limited to input values for a test case [30], we introduce a representation that focuses on execution traces (sequence of method calls and their I/Os) to better distinguish failing and passing tests.

---

Listing 1. A code snippet and its failing/passing test cases.

```
public double formula(int b, int e, int c){
        int p = power(b,e);
        if (c > 0)
                return p/c;
        else
                return -1;
}
public int power(int b, int e){
        int r = 1;
        if check(b,e)
                while(e>0){
                        r = r * b;
                        e = e - 1;
                }
        return r;
}
public int check(int b, int e){
        if(b==0)
                throw new Exception("Undefined");
        return 1;
}
//Tests
public void test1(){
        assertEquals(formula(2,0,3), 1/3);
}
public void test2(){
        assertEquals(formula(2,-1,3), -1);
}
```

Since our approach requires a training set that is collected from historical versions of the SUT to be able to represent a new test, it can only be applied if there are some existing tests for a project to start with. Thus, for the example in this section, we cannot run our model. However, Figure 1 visualizes a similar example from our dataset (Lang Project V61, in Defects4J), where the failed test's vector (in blue) is significantly different compared to other passed tests that are covering the same buggy code snippet [†]

The above example motivates how our embedding can distinguish the failing test case from the passing ones, which in turn suggests that tasks such as test generation and prioritization can leverage this information. In section 5, we will explore this idea in more details, in the context of test prioritization.

## 3  BACKGROUND

In this section, we provide background needed to better understand our approach and the baselines we will compare it to.

**Word Embedding and Sequence Learning:** Embedding is a mapping from an object to a vector of continuous

---

[†]The visualization is created using our proposed Test2Vec model to vectorize test case traces into an N-dimensional space followed with a tSNE [54] dimension reduction approach to reduce the dimensions into two (for the sake of visualization).

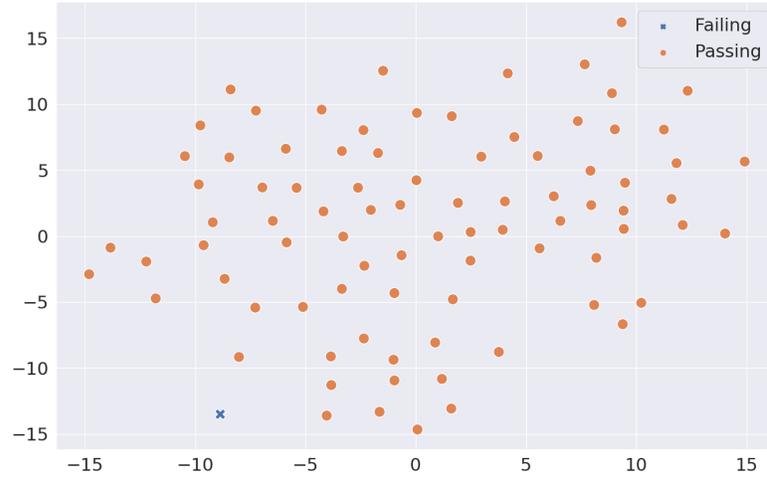

Fig. 1.  Test suite StringUtilsTest of Project Lang61 embedded by Test2Vec and visualized by tSNE. The axes are the latent features learnt by tSNE.

values/numbers. In the context of learning sequences of words, embedding is a mapping from each word to a vector of real numbers that can be used in mathematical models for tasks such as visualization and prediction. With recent progress in word embeddings, embedding models can learn about the semantic relation between words in a text and consequently can better represent each word, based on the context. Intuitively, the multiple dimensions of the vector[‡] allows the model to express different forms of semantic similarity in different "directions" of the (numerical) vector space.

When using an embedding, the distance between words in the text can be measured by calculating their vector representations distance. One of the most used distance metrics for embedding is the cosine distance. It is measured by the cosine of the angle between two vectors which does not rely on the magnitude. Euclidean distance is another alternative. However, in high dimensional spaces, the cosine function is more commonly used [14].

Early word embedding techniques such as Word2Vec by Mikolov *et al.* [56] employed neural network models for computing a continuous global vector as a representation for each word by preserving semantic relations in very large datasets. These approaches usually build a global vocabulary using unique words in the documents of the dataset. Using the other words it tends to appear close to, the representations for each word are then learned. Therefore, semantically similar words tend to be close to each other in the vector space, [57]. However, learning a global representation for each word means that these model ignore the contextual meaning (a word's meaning derived from its surroundings) of the words.

As an alternative, more recent embedding methods such as GPT [11] and BERT[18], known as contextual embedding methods, employ the whole or parts of a sentence to learn sequence-level semantics. Such techniques can learn different

---

representations for any word based on its context. In other words, contextual word embeddings capture its semantics in context, so even though it is the same word, it can be represented differently under different conditions.

**CodeBERT:** In recent years, large pre-trained models have shown promising results on various text analysis problems. BERT is a pre-trained model that achieved state-of-the-art performance on many NLP tasks [19]. It is a transformer-based neural architecture pre-trained on large texts with self-supervised objectives, Masked Language Model (MLM, predict which word was masked out) and Next Sentence Prediction (NSP, predict the sentence that follows). The self-supervision is very important since it allows the use of much larger datasets without manual labeling. BERT considers both the preceding and the following contexts to learn better contextual representations. The success of the BERT model also has led to BERT variants that are pre-trained on specialized corpora [9, 44]. CodeBERT by Feng *et al.* [28] is a BERT variant for both Natural Language (NL) and Programming Language (PL). It learns general-purpose contextual embeddings for both NL and PL. They evaluate CodeBERT on two downstream NL-PL tasks, natural language code search and code documentation generation. In this paper, we use CodeBERT as the state-of-the-art sequence learning model and apply it on our execution traces, as a baseline. Note that CodeBert is trained on static source code and not execution traces, but there is no other pre-trained large language model for execution traces, either. So using transfer learning we fine-tune CodeBert to be applicable on our trace sequences. The intuition is CodeBert will learn a representation based on the code-like nature of of the traces (including the method names and parameters etc.). Then the fine-tuning step provides a more accurate representation.

**Transfer Learning:** The concept of transfer learning refers to the reuse of a previously created model on a new problem [69]. Due to the fact that it is able to fine-tune and then use deep neural networks on problems where there is relatively little data, it is currently popular in deep learning. Essentially, the model is pre-trained on a large dataset and then fine-tuned by additional training on a much smaller dataset, for a specific task. Using these pre-trained models as a start point tends to improve results significantly compared to training models from scratch on the limited dataset. The pre-trained models are usually trained on general tasks, such as language modeling in NLP. As a result of this pre-training, the model will learn general-purpose knowledge that can then be "transferred" and used in downstream tasks.

**Test Case Prioritization:** Test Case Prioritization (TP) is the problem of ordering the test cases within the test suite of a system under test (SUT), with the goal of maximizing some criteria [74], e.g. executing failing test cases early. TP is mainly used in practice during regression testing and in the continuous integration setups, where new changes to the code should not break the existing features [38]. The traditional TP techniques focus on maximizing code coverage of some sort using an optimization technique such as a greedy algorithm [20, 45] (which is one of our baselines in this paper).

Another common TP category is fault-based or history-based TPs, where the test cases that have failed in the past (or are similar to the failed tests from history) are ranked higher [22, 62]. We have incorporated this idea as one the heuristics in our approach.

Finally, the last common category is diversifying test cases [26], which can be applied on different representations including the specification models [40], outputs [25], and execution traces [26, 59]. The main idea behind this category of TPs is that similar tests cover similar features and behaviour. Therefore, diversifying tests should result in a more even coverage of the feature/behaviour space. Diversity-based TP can also be seen as an anomaly detection, where the failing test's behaviour is considered as an atypical pattern. We have used this idea as the second heuristic in our approach.

In summary, our proposed TP combines the two common categories of TP heuristics and applies them on test cases' behaviour, represented by several encoding and embedding methods on top of program execution traces.

## 4   *TEST2VEC:* AN EXECUTION TRACE EMBEDDING

The key idea of this paper is a new embedding model for dynamic executions of test cases. The idea is that after applying our embedding on each test case execution (execution trace), that test's run-time behavior is mapped to and thus is represented as a vector in an N-dimensional latent space. These embedded vectors can then be used to analyze a test suite as well as individual, and groups of, test cases. For instance, using them one can predict whether a test case will pass or fail, using a simple softmax classifier and the actual, historical test outcomes (as target labels). Alternatively, a test case with an anomalous, i.e. different, behaviour within a test suite can be detected by calculating the distance between such vectors in the latent space (vectorized representation) of the test suite. Much like how embeddings have been used for natural language processing in the deep learning literature, test embeddings can be the basis for many different, downstream tasks in automated testing.

Thus, we propose "Test2Vec Embedding", a neural embedding model leveraging pre-trained models, to vectorize test execution traces. The inputs to our embedding model are traces (sequences of method calls, their input parameters, and their outputs). The trace embedding model is fine-tuned based on the corresponding test case's pass/fail label per execution trace. There are other options for fine-tuning the model, such as class name or method name prediction. However, since we are more interested in the failing behavior prediction, for our particular down-stream task (TP), we argue that using the pass/fail label as the final objective for fine-tuning makes more sense than the other options.

The specific Test2Vec implementation we propose and evaluate here uses CodeBERT layers that are originally pre-trained on generic tasks, such as masked label prediction, on both programming and natural language data. CodeBERT's job in Test2Vec is to embed methods and I/O parameter names, individually, which is the closest sub-task within trace embedding that can leverage the learnt representation from CodeBert's pre-trained model on static source code. However, to be useful for execution trace embedding (vs. the original static code embedding) these layers are then fine-tuned, end-to-end, while we are training the Test2Vec model for test case prioritization that let the model learn the passing/failing behaviors for the SUT. In this study, our fine tuning labels are pass/fail labels that are collected by Junit execution of the test cases, but in general, our Test2Vec method can be applied on variety of testing-related tasks with potentially a different fine-tuning.

Our hypothesis (which we will verify by the experiments results) is that a project's dataset including the current revision's test cases as well as historical testing results from past revisions, will be enough for fine-tuning the model. The assumption is that these trained models will help us with e.g., predicting which test cases are more likely to fail by classifying their traces based on the historical failures of similar traces in the past. It will also help analyzing an individual test case's embedding, in relation to other test cases in the current test suite. For instance, the embedding vectors of similar test cases within a class, which are all covering the core functionality and have a lot of overlap, should become closer to each other. Whereas, the vector representation of a test case covering the less tested feature, within the class, becomes more like an outlier in the test suite (more distant from the centroid). In addition, anomalous run-time behaviour resulting from executing a failed test case will also be likely represented as anomalous vector, within the behaviour space (set of vectors in a test suite). However, critical in this is that the sequences of methods called might not be enough to capture fine-grained differences in behavior. Modeling also input and output values can help in better distinguishing test case behavior.

The definition of an execution trace in this study is, thus, a sequence of method names (all methods of the class under test that are invoked during execution of the given test case) with their inputs and outputs. The choice of what to include in the trace is somewhat arbitrary and a systematic study is needed in the future to identify the elements that contributes the most and thus should be included, per task. However, the method calls, their inputs, and theirs outputs seems like the minimum that are suggested in various literature [27, 39, 77].

We call each triplet of $< Outputs, MethodName, Inputs >$ a *Context*, since each method call is, in fact, a context for other method calls within the whole sequence. Given that each trace can have a varying number of method invocations (Contexts), a challenge is to map all of these Contexts into one single embedding vector.

Figure 5 and Figure 4 shows an overview of our approach, which consists of three main phases, as follows:

**Phase1: Test case execution and trace collection:** The process starts with running all existing test cases per project (in our case, the developer-written unit tests) for the current version of the SUT and its historical versions. That means our dataset for each project is based on the current and historical test cases within the same project. Historical test cases are each run on their corresponding source code versions (since most older versions have obsolete test cases that are not executable or even compilable, in the most recent version of the program). This will generate traces for all developer-written test cases including failed test cases. However, since normally there are many more passing tests than failing ones, this will lead to a highly unbalanced dataset with hundreds of passing traces and a few failing traces. To mitigate this problem, we use mutation testing to automatically seed artificial faults and generate more failing test cases.

To log the execution traces of the test cases, we have instrumented the SUT code that allow us to capture the method calls (including nested calls when the SUT is called) and all I/O names, types, and values at run-time.

Table 1. List of mutation operators used by Major framework to mutate PUTs.

| Mutation operator | example |
|---|---|
| Arithmetic Operator Replacement | a*b → a/b |
| Logical Operator Replacement | a&b → a\|b |
| Conditional Operator Replacement | a&&b → a\|\|b |
| Relational Operator Replacement | a<=b → a<b |
| Shift Operator Replacement | a»b → a«b |
| Operator Replacement Unary | -a → !a |
| Expression Value Replacement | a=b → a=0 |
| Literal Value Replacement | true → false |
| Statement Deletion | break → <no-op> |

As mentioned, each execution trace includes the sequence of invoked method calls while running the test case. This means we are not limited to the method names which are in the test script and can include all nested calls, invoked at run time. However, to keep each trace focused on the Class Under Test (CUT), if a method of the CUT invokes methods belonging to another class, only the first method invocation from the external class is included (the method that was called from CUT) in the trace and the rest of calls are ignored. Essentially in this way we isolate units–CUTs– and focus on unit testing rather than integration testing.

The execution trace also includes the method's input parameters (if any), outputs, and their data types. Therefore, our traces may contain a lot of numerical data (alongside textual data). Research [78] have shown that most NLP techniques cannot work well with numerical data, especially large values. They also demonstrated that the state-of-art NLP models, such as BERT, that rely on sub-word tokenizers have difficulty embedding large numeric and float values. For this reason, we tried a simple abstraction technique for primitive types that replace the value of each primitive parameter with one of the predefined special token, based on its value. For $int, short, long, double, float$, we simply break the range into five categories: "negative big value (NBV)", "negative value (NV)", "zero value (ZV)", "positive value (PV)", and "positive big value (PBV)". For $string$ and $array$, we just abstract into empty vs non-empty string or array. More details about these tokens can be found in the replication package [7]. For non-primitive parameters (objects), we include their object type and their reference value. Please note that this approach for abstraction is only one plausible way and there is still room for more research in this area, to determine how to best handle numerical data and complex objects within execution trace embedding, which is in our future work.

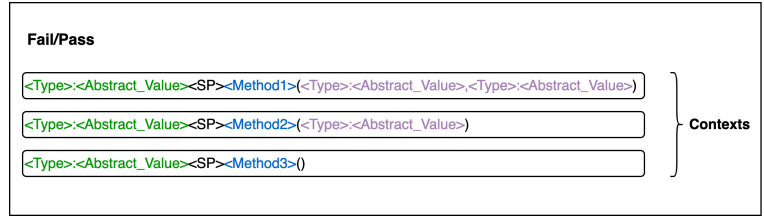

Fig. 2. Trace template after pre-processing.

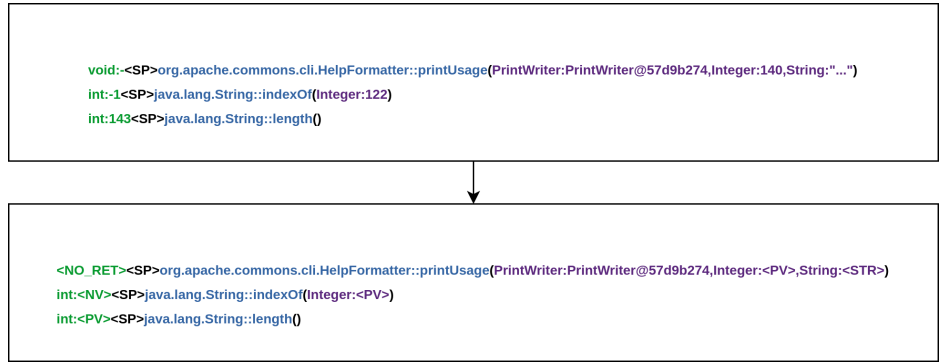

Fig. 3. A real example of execution traces passed to Test2Vec, before and after preprocessing

Figure 2 shows the template of the collected traces after prepossessing, which includes a sequence of tuples following this format: *Test Result, Output Type, Output Value, Method Name, Parameter Type, Input Value*, in which *Test Result* shows whether the test case has been passed or failed.

Figure 3 shows an example of raw execution trace and its representation using the template.

In practice, the execution traces typically vary quite significantly in length. To have a consistent, limited size, and less sparse training data, we have limited the number of contexts in a trace (number of calls) to MAX Context. Besides, the CodeBERT model accepts a limited number of tokens (in our study, we used the maximum number of tokens, 512 tokens, which it can process). Therefore we limited the number of contexts for each trace to 128, by excluding the contexts from the middle of each long sequence, since these contexts for long trace sequences, tend to be repeated calls in loops. This value is such that most of the traces will have less than 512 tokens for each CodeBERT model, and more than 75% of traces remains unchanged.

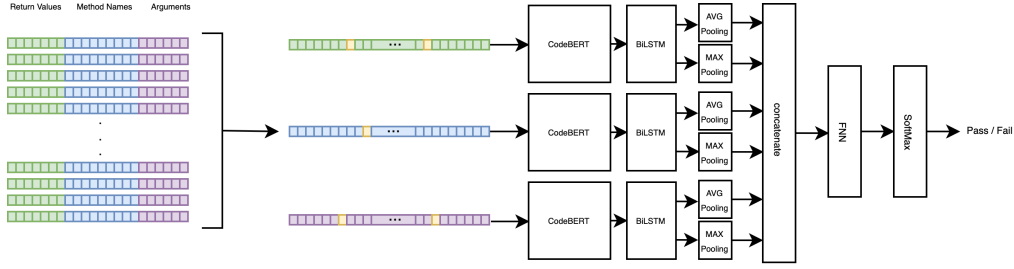

Fig. 4. Test2Vec Overall Model Architecture.

**Phase2: Trace embedding generation:** After extracting all method calls from each trace (preserving their order), we create three different sequences for outputs, method names, and input arguments (we use <NO_ARG> in cases where there is no Input or Output). We pass each sequence to a distinct CodeBERT model to generate a separate embedding, per token, in each sequence. To fine tune each CodeBERT, we start with the weights of the pre-trained transformer models in CodeBERT and then fine-tune the models on our training data (traces from 20 versions of a project), while classifying the traces as pass or fail.

We use a small LSTM layer after each Transformer, which is a very common way to extract sentence embeddings from BERT models [60]. At this point, we have a separate embedding for each token in each sequence. We then use Pooling layers **??** to generate the final representation for a sequence and aggregate these token embeddings. We use both Max and Average Pooling layers and concatenate those to generate a final embedding for the sequence.

In the end, we concatenate the three embeddings that we calculate for the sequences and use this as the embedding for the trace.

**Phase3: Training the classifier and fine-tuning the embedding models :** After generating a single representation vector for a test case, called *Test Vector*, we use this representation alongside the label of the trace, the passing or failing label that we collect in Phase 1, to both train the end-to-end model (including fine-tuning the CodeBERT layers). In this way, we can have embeddings that capture more information regarding the fail/pass behavior of each trace and a classifier that predicts the probability of being fail or pass for each embedding (learn a conditional distribution: $P(Failing|Embedding)$).

The training of the model is done in two steps. We, first, freeze the embedding generation layers. This is necessary because without freezing these layers, weights of these layers will be updated from the first epochs, and since the classification layers are not accurate at the beginning of the training process, the pre-trained weights start to update randomly, and we will lose the benefits of transfer learning. Therefore, we started the training process by freezing the CodeBERT layers and only training the rest of model.

After training the LSTM and the classifier layers, we start the CodeBERT fine-tuning process. For fine-tuning, we start with unfreezing the embedding layers and reducing the learning rate of the model. This will help the model update weights for the embedding layer without altering them too much and forgetting the knowledge it learned during the pre-training process. We then resume the training for a maximum of ten more epochs. This will let us update the weights of the whole model end-to-end.

After these three phases, Test2Vec provides us with a representation per test case that is fine-tuned based on historical failing test cases. In the next section, we evaluate this representation for a test prioritization task.

## 5 EVALUATION

In this section, we describe the details of the experimental evaluation including the specific research questions, the design of the experiments, as well as the findings.

### 5.1 Hypotheses and Research Questions

We investigate two hypotheses in two main research questions as follows:

**RQ1) How effective is Test2Vec embedding for test prioritization, when the tests are ranked using a classifier trained on historical data?**

The first hypothesis of this study is that Test2Vec allows a fine-grained representation of test cases which can improve performance on downstream tasks, with respect to failing/passing behavior, compared to code coverage and state-of-art code and trace embedding techniques. To validate this hypothesis, in RQ1, we compare Test2Vec with alternative representation methods for a specific downstream task of test case prioritization (TP). We argue this is both an important, well-studied, and representative task in automated testing. Our default approach (Test2Vec classifier-TP) for test prioritization based on Test2Vec is using the classifier (softmax) layer's probabilities that we trained for the pass/fail label prediction task. To address RQ1, specifically, we answer these two sub-RQs:

*RQ1-1) How effective is Test2Vec classifier-TP compared to a code coverage-based prioritization (state of the practice)?* The goal of this sub-RQ is to justify the whole idea of test case behaviour representation compared to simple code coverage, for a task such as TP. As a basic coverage-based TP, we use the Greedy-based additional coverage TP [45], which ranks the test case that covers the most uncovered lines/branches higher, at each step. This baseline has been used in much of the TP literature as coverage-based baselines. Our proposed approach will be Test2Vec classifier-TP that ranks test cases using a softmax classifier trained on historical data. This classifier calculates the probability of being a failing test for each trace that is used for ranking test cases (simply by sorting the traces descending based on their failure probability).

*RQ1-2) How effective is Test2Vec classifier-TP compared to basic and state-of-art code and execution trace embedding techniques from literature?* The goal of this sub-RQ is to justify the use of our proposed customized and relatively advanced embedding compared to existing code- and test-based embeddings. To do so, we compare Test2Vec classifier-TP with test case prioritizations calculated with classifiers that use (a) one-hot-encoding as a simpler embedding, (b) CodeBERT, as the state-of-art code embedding model, and (c) an LSTM-based model, as the state-of-art execution trace embedding model.

**RQ2) Which of the two prioritization heuristics (similarity to past failing tests or test diversification) is better to be used with Test2Vec for a test case diversification task?**

Our second hypothesis is that the representational power of the test case embeddings we propose can be leveraged with diversification as a natural, performant, and potentially complementary heuristic for the TP problem. Since diversity has shown promise in prior test automation and prioritization work [26, 42? ] a rich representation of execution traces might benefit such prioritization methods. To leverage this heuristic, we used an algorithm that ranks a set test cases iteratively, by prioritizing the most diverse test case (i.e., the most dissimilar test case compared to others in the latent, embedding space), in each iteration. While this is not the most efficient way this can be done our main goal is in evaluating the power of the embeddings; future work can investigate more efficient approaches. In order to examine the effectiveness of this heuristic, we designed the following sub-RQs:

*RQ2-1) How effective is test diversification using Test2Vec embeddings (Test2Vec diversification-TP), for test prioritization?* In this sub-RQ, we compare the results of our default TP (Test2Vec classifier-TP) from above with a diversification-based one (Test2Vec diversification-TP). Given that the two heuristics potentially target different types of fault, we also consider if the approaches can be combined by a classifier that selects which method (classifier or diversification) should be used. We call this method Test2Vec combined-TP.

*RQ2-2) Can combining diversification and history-based classification (Test2Vec combined-TP) outperform each of them individually?* The hypothesis of this sub-RQ is that the bugs that are similar to historical failures can be better ranked by default TP and those that are new and pose as anomalies (within the existing tests) can be better ranked by diversification-TP. We also argue that if this is the case it adds further support that the embeddings we propose can capture rich information about test case behavior. Thus, it makes sense to propose a combined TP that dynamically figures out which approach is better suited per test suite. Therefore, we propose "Test2Vec combined-TP" approach, which is a classifier that is trained on the historical version of the project to predict the better heuristic per test suite, given some features from test vectors and their labels. RQ2-2 compares Test2Vec combined-TP with its constituent alternatives when used individually.

## 5.2 Evaluation Metrics

Similar to representation learning literature in NLP [72], evaluation of a representation method is more meaningful through a downstream task. In our case, we chose TP as our down-stream task given that it is one of the main use cases of automation in real-world software testing and has been well studied in the literature. Another reason is that unlike for automated test generation, which depends on multiple factors (e.g., the abstract test representation and design, the input data generation strategy, and approaches to make executable tests out of abstract tests, etc.), automated TP depends heavily on fewer factors, i.e. the test representation and the ranking strategy. Thus, we vary both of the latter in our experiments.

The evaluation metrics we have used are standard metrics from the TP domain, such as APFD [66] and the rank of the first failing test [12] (FFR). The FFR metric represents the normalized position of the first failed test case for a test case prioritization method. Equation 1 defines FFR more formally.

$$FFR = 100 * \left( \frac{TF_1}{n} \right) \qquad (1)$$

where n denotes the number of test cases to be prioritized and $TF_1$ is the number of tests which must be executed before first fault is detected by the prioritized tests. Note that we normalize the ranks based on the number of test cases, since otherwise a TP's effectiveness in ranking test suites A and B would be the same, when it ranks their failing tests as #5 in both A and B, even if in A there is only 10 test cases (the TP found the failed test only after running half of the test suite) and in B there are 100 tests (the failed test was detected among top 5%).

Our second metric, the average percentage of fault detection (APFD), captures the average percentage of faults detected by a set of prioritized test cases. It can also be seen as the Area Under Curve where the x-axis is the ranked test cases and y-axis is the cumulative number of faults detected, when the tests are executed following the x-axis order. The higher the APFD, which ranges from 0 to 100, the better the fault detection system. Equation 2 defines APFD more formally:

$$APFD = 100 * \left(1 - \frac{TF_1 + TF_2 + ... + TF_m}{nm} + \frac{1}{2n}\right) \qquad (2)$$

where n denotes the number of test cases to prioritize, m is the total number of faults detectable by these test cases, and $TF_i$ is the number of tests which must be executed before fault i is detected. APFD is more informative if there are multiple choices of failing test cases and faults to prioritize so that the area under curve is a good "summary" of the prioritization effectiveness. Otherwise, if for instance, there is only one bug and one failing test, simply measuring the rank of failing test is a more direct and meaningful metric.

Because of the characteristic of our dataset (Defect4J), which mostly has only one failing test per bug and only a few bugs per version of the project, APFD is not suitable on Defect4J's real bugs. Therefore, we only use the rank of first failing test as our evaluation metric when dealing with the real bugs. However, we also injected seeded faults using mutation testing into the projects to be able to compare techniques with respect to ranking seeded bugs. Since we have many failing test case in this case, we can use APFD to compare models.

Beside the choice of metrics (FFR vs APFD) another design choice we have is at which level to apply such metrics (i.e., whether to prioritize test cases of one class or the entire project's–all test cases of a version combined). The two approaches can be debated. Applying TPs on one class is more accurate and provides better rankings since we only compare behaviorally similar test cases. Comparing all test cases from all classes are sometimes very noisy. However, to begin the TP process in the class-level, one must know which class to focus on.

To deal with this challenge we evaluate our approach in both levels. In the whole version-level, we use APFD to summarize the effectiveness of the TP over all test cases and faults (i.e., both real and the seeded faults). To evaluate the TPs at the class-level, we use the FFR (normalized rank of real faults). In this way, we get a sense of TPs' effectiveness at both levels. However, our main metric is still the FFR on the class-level for the following reason: (a) It is the most direct metric for ranking, in particular when there is only one fault to catch. (b) We use it on the real faults. (c) As we see later in this section, APFD on mutants does not work on our diversity heuristic. (d) In practice, there are other ways to decide which source classes should be tested (e.g., using defect prediction, or simply based on last changes to the code) and thus a TP technique can know in advance which test suite(s) to target.

So in summary, whenever we report AFPD results they are the summary of TP's effectiveness on ranking all test cases in all test suites, per version, to detect mutants. Whereas, reported FFRs are the results within the failing test suite (only test cases of one class), with respect to catching real bugs.

### 5.3 Dataset and the Data Extraction Procedure

In this study, we run our experiments on 10 open-source Java projects from the Defects4J project [48]: Commons-Lang (Lang), Joda-Time (Time), Commons-Cli (Cli), JFreeChart (Chart), Commons-Compress (Compress), Jsoup, Mockito, jackson-databind (JacksonDatabind), jackson-core (JacksonCore), and commons-math (Math). Defects4j provides multiple revisions per project, each one providing a faulty and a fixed source code version. For each project, we used the last 25 revisions (20 older revisions for training and 5 most recent revisions for testing). Table 2 shows the number of versions and classes per project.

In Defects4J, each faulty version has exactly one bug (failure) and one corresponding failing test suite (called "triggering test suite"), with at least one failed test case (called "failed test"). With respect to seeded faults, as it is the norm for mutation testing, we make many mutants per version, where each mutant isolates one faults. In most versions of the projects, there is only one failing test case in the failing test suite, but there are also versions that have multiple failing test cases.

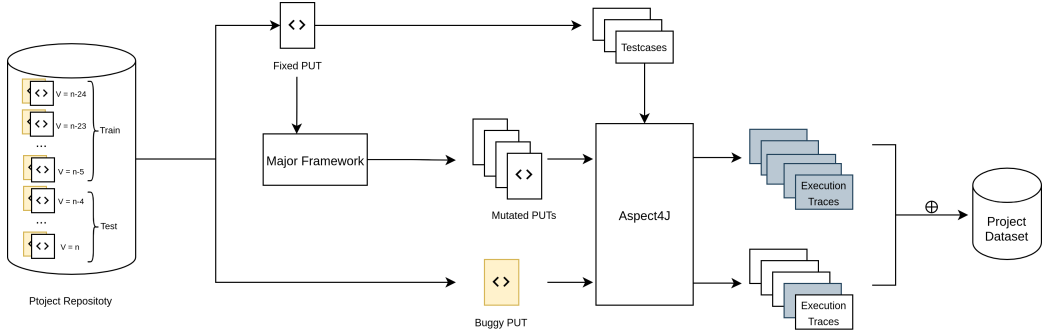

Fig. 5. Overall process of generating the trace datasets, per project. "failed" traces (shown as colored boxes) are collected from running tests on mutants or real bugs.

Figure 5 shows the process of generating traces for each project in Defect4J. First, for each version of the project, we fetch its buggy version and run all its corresponding test cases. In our dataset, this will generate a set of traces that are mostly passing traces and only a few failing test cases (usually one or two failing test cases). To make the dataset balanced, we then fetch the fixed version of the program and give it to our mutant generation tool, the Major [47] framework. The Major framework will generate a great number of mutated versions of the source codes that are used as the Program Under Tests (PUTs). For mutation generation, we used the common mutation operators (the list of the operators we used in this experiment can be find in the Table 1). We run the test cases on the PUTs and collect traces for only the failing test cases. We repeat this procedure, by randomly selecting a new PUT and running the test cases on, until we have the same number of failing and passing traces for each version of the program. To avoid the problem of equivalent mutants, we ignore PUTs with no failing test cases.

As discussed, we augment our dataset with the failing traces on seeded faults, until we have an exact balanced dataset. Given that the total number of failing tests on seeded faults are more than needed (to keep the balance between pass and fail) we have to select a subset of failing traces. Therefore, we randomly select only one failing trace per seeded fault, so each trace covers a unique fault.

Since the method calls related to assertions or exceptions may contain information about test case results in their input parameters or outputs, we remove all the method calls related to exceptions and assertions from the traces in order to prevent data leakage.

We ran all developer-written test cases of the projects and aggregated all traces from all faulty versions and mutated versions of a project, to create a dataset per project. It should be noted that some of the tests might be repeated across versions of a project. However, since the tested source code of different versions is not always the same, we kept the repeated tests as their traces might still differ.

At the end, we had a total of 192,768 execution traces over the 10 projects, ranging from around 2,186 to 82,384 traces, per project. Table 2 reports the number of traces per project as well as the distribution of traces in the failing test suite, per project and failed class.

We used Train and Validation datasets for training and evaluating the Test2Vec embedding model, and the Test dataset for analysis of the downstream task (TP). Thus, each project has its separate dataset and trained model. We then split each project's trace dataset into Train, Validation, and Test sets. Among the last 25 included version per project, we use the last five versions (21-25; where V25 is the most recent version) as the Test set. For the other 20 versions, we aggregate all the traces, including traces related to mutated source codes. We, then, split it into Training (80%) and Validation (20%) sets.

Table 2. Projects Under Study.

| Project | #Total Traces | #Trace Per Failed Class | | |
|---|---|---|---|---|
| | | Min | Median | Max |
| Chart | 82384 | 57 | 76 | 1192 |
| Cli | 2994 | 18 | 23 | 170 |
| Jsoup | 6798 | 6 | 23 | 132 |
| Compress | 8352 | 7 | 11.5 | 17 |
| Lang | 15732 | 11 | 19 | 73 |
| Time | 38038 | 54 | 104 | 153 |
| JacksonCore | 5590 | 7 | 8 | 16 |
| JacksonDatabind | 21478 | 15 | 15 | 18 |
| Mokito | 2186 | 7 | 8 | 9 |
| Math | 9216 | 7 | 11 | 16 |
| Total | 192,768 | | | |

### 5.4 Design

To answer **RQ1-1**, we implemented two TP methods: (a) an Additional Greedy Algorithm [38] that maximizes a given coverage (in our study line and branch coverage) and (b) a Test2Vec classifier-TP Algorithm that ranks traces based on the $P(failing|trace)$ calculated by a softmax classifier that ultimately shows the probability of a test case failure based on the similarity of its trace to historical failures.

**Coverage-based TP:** Since our focus is on the benefits of our proposed embedding approach, rather than on the optimization step of TP, we use a Greedy optimization algorithm for both TP methods, mentioned above. Future work can replace the Greedy algorithm with alternatives like meta-heuristic methods.

The additional Greedy Algorithm, applied on line (branch) coverage, simply starts with the test case with the highest line (branch) coverage. Then, the next test case that covers the highest "uncovered" lines (branches) will be added. This will continue until all test cases are ordered. Whenever there is more than one test with the same best coverage, one is chosen randomly. JaCoCo was used to collect the line and branch coverage values per test case.

**Test2Vec classifier-TP:** To calculate Test2Vec classifier-TP, first, we train the "Test2Vec model" for each project using its training dataset (which includes sequences of method calls, their outputs, and their input values). Since we train the Test2Vec model for classifying execution traces to fail or pass, we can use the same probability that is calculated by the softmax layer for ranking the trace executions, as well. In other words, these probabilities show the confidence of the model about an execution trace belonging to a failing test case and thus being prioritized in the TP process.

The goal of **RQ1-2** is to compare Test2Vec with the current body of research [28, 77] in execution trace and code embedding techniques, for the TP task.

Our first baseline in this sub-RQ is a basic One-Hot encoding. In this approach, we simply extract all the method calls (that are called from the test case and class under test) from the trace and create a one-hot representation of these method calls. Since there are many possibilities for the input parameters and output values, it is not possible to use one-hot encoding to represent them. Therefore, we ignore input parameters and outputs in this approach. The output of this technique is a sparse diverse encoding for each trace. We then pass these encodings to a classifier to be classified as pass or fail. The main reason for this baseline is to understand whether including I/O and consequently being forced to use advanced embedding is necessary or a simple one-hot representation is enough for execution trace encoding for TP.

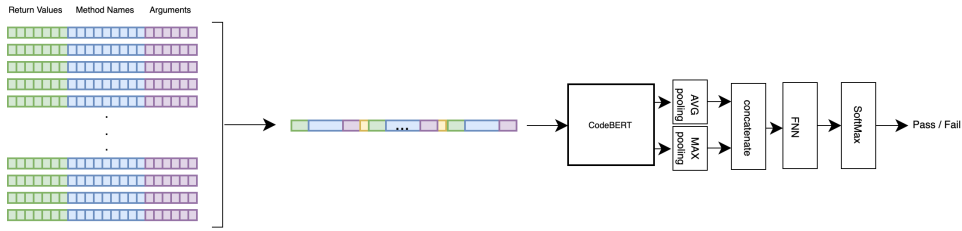

Fig. 6. CodeBERT-TP Overall Model Architecture.

Our second baseline in this sub-RQ is the most related work (we call it the LSTM-based trace embedding) in the domain of execution trace representation for testing [77]. LSTM-based trace embedding uses encoded arguments and return values and concatenates these representations along with one-hot representations of the caller and callee methods to generate the same size vectors for each method call in the trace. These vectors then will be passed to an LSTM and MLP (Multi Layer Perception) to be classified as pass or fail.

The last baseline in this sub-RQ is the state-of-the-art neural program embedding model, **CodeBERT**. The idea here is we use one CodeBERT to embed the entire execution trace and do not bother with the customized architecture of Test2Vec (which uses CodeBERT but adds extra structure and layers to it; See Figure 4). Since CodeBERT is a generic embedding model, we need to at least fine-tune the model for the TP task. To do so, we take the final fixed-size embeddings of each trace from the CodeBERT using maximum and average pooling strategies and concatenate these vectors together (see Figure 6). These embeddings are then passed to an MLP layer and a softmax classifier to predict the final label.

For the actual test prioritization in both LSTM-based and CodeBERT TPs, we apply the very same process as Test2Vec classifier-TP's, i.e., ranking tests based on their softmax classifier's probabilities.

In **RQ2-1**, we propose Test2Vec diversification-TP that ranks traces based on their distance to the centroid of the test suite and compared it with the Test2Vec classifier-TP.

**Diversification-TP:** To calculate the ranks based on Test2Vec diversification-TP, we take all embedded vectors (the same ones from Test2Vec classifier-TP) per test suite and compute the cosine distance [15] of each test vector to the center of all test vectors in the test suite. The TP algorithm then ranks test cases based on their distances descendingly.

In **RQ2-2**, we design the combined-TP approach. Recall that our hypothesis is that when the failing test case is similar to the historical failures the classifier-TP usually works better. Whereas, when the failing test is not close to historical failures, and it is more like an anomaly in the test suite, the diversification-TP works better. Therefore, the proposed combined-TP approach merges the two heuristics, by deciding which one to use per test suite. It uses a logistic regression model as a classifier to decide which technique should be used for detecting the failing test case. This classifier uses two features for prediction. The first feature is the average probability calculated by the softmax model for the top 5 test cases in the ranking. The second feature is the average distance of the top 5 most outlying points to the center of the test suite divided by the average distance of all points to the center.

Intuitively speaking, the combined-TP approach checks the top N tests ranked by each heuristic and chooses the heuristic which is more confident in its ranking predictions (either through probabilities or distances). The reason to use the probabilities and distances of top N tests rather than all test cases, is that there is only a few failing tests (mostly just one) per test suite. Therefore, the ranks of the rest of test cases is not predictive of the effectiveness of the classifier-TP vs. diversification-TP, and acts more like noise in combined-TP's decision. The actual value of N is a hyper-parameter for classifier-TP and N=4 was chosen here based on our small tuning experiments with N= 1 to 10.

In terms of evaluating the TPs in each RQ, we have different approaches for RQ1 and 2. In RQ1, we use both FFR and APFD.

Regarding the FFR analysis, all the coverage-based, classifier, and diversification TPs are applied to the failed version of each revision. To have a statistically and practically meaningful TP analysis using our metrics, we have only included those failing test suites, from the test dataset, that have more than 5 test cases, per buggy method. This results in excluding 8 out of 50 failures (one version from Cli, one version from Compress, two versions from JacksonCore, one version from JacksonDatabind, and three versions from Mokito).

Regarding the APFD analysis, the traces are collected from the mutants as explained in 5.2) so there is no concern about the limited number of data point, as in the FFR analysis.

In RQ2, we only use FFR. The reason is that diversification-TP works based on detecting a few outliers among mostly typical test cases in a test suite. That's why if we seed hundreds of fake faults, by mutation to measure APFD, the bugs can not all be correctly identified as anomalies.

In both RQs, the TPs that have a random component (coverage-based-TP techniques, since there are many test cases with the same coverage score, where we had to break the tie randomly) are executed 30 times per version and we record the median values (AFPD and FFR) of those 30 runs for that version. To compare two TPs in any RQ, we look at the mean, median, and statistical significant test results of FFRs and APFDs over the 50 (5X10) project versions. Therefore, depending on the TP, each of these 50 values per distribution is either an individual metric (AFPD or FFR) or a median of 30 measurement of that metric.

We use a paired non-parametric statistical significance test (Wilcoxon signed rank test [68] with p-value less than 0.05) and calculated the effect size (rank-biserial correlation [68]), every time we compare two distributions of ranks over their 50 samples.

### 5.5 Configurations & Environments

We compile and run projects with *Java1.7*, as required by the versions in Defects4J. Our code is implemented in *Python3.9.7*. For model training, we have used *TensorFlow 2.7.0* and *Keras* libraries for implementing our neural network model. We, also, have used the CodeBERT implementation from the *huggingface/transformers* framework and loaded the pre-trained CodeBERT-base weights. The test vector size (i.e., number of dimensions in the latent space) is set to 100 where each vector generated from each CodeBERT model have size of 768. This configuration was found after some ad-hoc tuning but it might not be optimal and it can be easily set to any size in the configuration file. Also, the models were trained to 40 epochs, and each CodeBERT model fine-tuned for 10 epochs. Training and evaluation of the deep learning model was done on a single node running Ubuntu 18.04, using 32 CPUs, 250G memory, and a Tesla V100 GPU.

The time that we consumed for fine-tuning the CodeBERT model varied depending on the size of the project. For example, it took about 26 hours to fine-tune the CodeBERT model in project Lang. In comparison, the training model for Test2Vec was around 37 hours. Note that training jobs are not frequent (once per project) and the inference times (vectorizing the test suite of a version based on the trained model) are negligible (a few seconds). Finally, prioritization time differences between different TP methods are also practically negligible (on average less than a second) and literally the same for all diversification and classifier-TP techniques.

Table 3. Statistics on the normalized ranks (FFRs) for RQ1, over all 42 versions under study, grouped by projects. The FFRs of coverage based TPs per version are the median of 30 runs. All median and AVG data are across the versions per project. The bold values are the dominating results.

| Project | Test2Vec | | Line Coverage | | Branch Coverage | | One-Hot | | LSTM-based | | CodeBERT | |
|---|---|---|---|---|---|---|---|---|---|---|---|---|
| | Median | AVG | Median | AVG | Median | AVG | Median | AVG | Median | AVG | Median | AVG |
| Chart | **5.70** | **7.73** | 25.00 | 31.67 | 56.67 | 43.39 | 18.29 | 17.37 | 26.89 | 25.22 | 12.28 | 15.12 |
| Cli | **12.87** | **13.70** | 36.37 | 41.89 | 44.65 | 43.14 | 29.79 | 30.04 | 34.65 | 30.58 | 17.42 | 18.05 |
| Jsoup | **13.04** | **16.27** | 18.18 | 21.89 | 46.81 | 44.94 | 27.27 | 33.68 | 24.13 | 35.27 | 30.43 | 28.50 |
| Compress | **14.28** | **14.67** | 69.04 | 54.53 | 48.82 | 49.21 | 15.96 | 21.82 | 34.82 | 38.20 | 20.53 | 23.92 |
| Lang | **10.52** | **12.00** | 45.61 | 57.22 | 50.90 | 47.34 | 27.28 | 35.41 | 31.57 | 27.61 | 18.18 | 22.39 |
| Time | **9.25** | **10.69** | 20.54 | 21.87 | 20.64 | 25.29 | 24.07 | 22.51 | 16.34 | 19.16 | 14.42 | 14.72 |
| Mokito | **18.25** | **18.25** | 73.00 | 73.00 | 25.78 | 25.78 | 26.98 | 26.98 | 36.50 | 36.50 | 25.39 | 25.39 |
| Math | **12.5** | **15.89** | 48.125 | 48.66 | 47.29 | 47.09 | 31.25 | 28.50 | 28.57 | 33.87 | 22.22 | 22.99 |
| JacksonDatabind | **8.88** | **9.44** | 60.00 | 56.38 | 56.57 | 66.16 | 37.50 | 41.18 | 36.66 | 32.50 | 24.44 | 28.88 |
| JacksonCore | **12.50** | **13.09** | 28.57 | 24.10 | 34.28 | 30.52 | 20.28 | 22.60 | 31.25 | 44.94 | 18.75 | 19.34 |
| Total | **12.50** | **12.86** | 36.93 | 41.33 | 47.05 | 43.27 | 27.50 | 27.27 | 28.57 | 31.39 | 19.09 | 21.69 |

Table 4. Statistics on the normalized ranks (FFRs) for RQ2, over all 42 versions under study, grouped by projects. The FFRs of coverage based TPs per version are the median of 30 runs. All median and AVG data are across the versions per project. The bold values are the dominating results.

| Project | Test2Vec | | Diversification-TP | | combined-TP | |
|---|---|---|---|---|---|---|
| | Median | AVG | Median | AVG | Median | AVG |
| Chart | **5.70** | **7.73** | 17.36 | 18.68 | **5.70** | 8.28 |
| Cli | 12.87 | 13.70 | 16.44 | 18.29 | **10.79** | **11.62** |
| Jsoup | **13.04** | 16.27 | 20.68 | 23.33 | **13.04** | **14.32** |
| Compress | **14.28** | **14.67** | **14.28** | 18.99 | **14.28** | 18.99 |
| Lang | 10.52 | 12.00 | 10.52 | 14.75 | **9.09** | **10.18** |
| Time | **9.25** | 10.69 | 11.92 | 12.12 | **9.25** | **10.31** |
| Mokito | 18.25 | 18.25 | **12.69** | **12.69** | **12.69** | **12.69** |
| Math | **12.5** | **15.89** | 19.54 | 31.01 | 14.94 | 28.44 |
| JacksonDatabind | **8.88** | **9.44** | 25.55 | 26.11 | **8.88** | **9.44** |
| JacksonCore | **12.50** | **13.09** | 28.57 | 24.10 | **12.50** | 17.85 |
| Total | 12.50 | 12.86 | 15.04 | 19.03 | **11.11** | **12.57** |

## 5.6 Results

In this section, we report and discuss the results of our experiments, per RQ.

**RQ1) How effective is Test2Vec embedding for test prioritization when the tests are ranked using a classifier trained on historical data?**

To address this RQ, we compared Test2Vec classifier-TP with baselines in two categories: basic traditional TPs (RQ1-1) and alternative embedding (RQ1-2).

**RQ1-1) Comparing with Coverage-based TPs:**

As shown in Table 3, among all 42 (50 - the 8 versions that did not have enough tests in the failing test suite) faulty versions under study, the normalized rank of the first failing test (FFR) in the original faulty versions for the Test2Vec classifier-TP has an average of 12.86 and a median of 12.50. In comparison, the line coverage(LC)-based TP has an average of 41.33 and a median of 36.93, and the branch coverage(BC)-based TP technique has an average of 43.27 and a median of 47.05. The relative improvement of the Test2Vec over the coverage-based TPs are: 68.89% (average) and 66.15% (median) over LC-based TP; and 70.28% (average) and 73.43% (median) over BC-based TP.

Table 5. Statistics on the APFDs, over all 50 versions under study, grouped by projects. The APFDs of coverage based TPs per version are the median of 30 runs. All median and AVG data are across the 5 versions per project. The bold values are the best results.

| Project | Test2Vec | | Line Coverage | | Branch Coverage | | One-Hot | | LSTM-based | | CodeBERT | |
|---|---|---|---|---|---|---|---|---|---|---|---|---|
| | Median | AVG | Median | AVG | Median | AVG | Median | AVG | Median | AVG | Median | AVG |
| Chart | **92.71** | **94.37** | 69.85 | 71.21 | 73.02 | 73.45 | 79.99 | 80.16 | 76.93 | 78.82 | 90.15 | 91.21 |
| Cli | **93.07** | **91.32** | 71.79 | 73.89 | 69.17 | 68.22 | 76.23 | 78.93 | 81.19 | 78.21 | 88.31 | 87.74 |
| Jsoup | **90.81** | **90.57** | 68.96 | 70.83 | 70.54 | 69.92 | 75.93 | 74.17 | 76.93 | 77.23 | 90.17 | 90.44 |
| Compress | **89.76** | **89.13** | 75.01 | 74.21 | 69.67 | 70.61 | 77.30 | 79.36 | 81.27 | 81.09 | 94.43 | 88.53 |
| Lang | **89.54** | **88.74** | 72.31 | 72.18 | 71.89 | 72.79 | 76.41 | 75.49 | 72.94 | 73.62 | 82.19 | 86.49 |
| Time | **92.35** | **91.70** | 74.82 | 76.53 | 73.47 | 72.28 | 76.64 | 78.92 | 75.07 | 74.32 | 85.92 | 86.30 |
| Mokito | **89.17** | **89.83** | 69.58 | 69.09 | 69.95 | 70.53 | 81.87 | 80.01 | 80.11 | 80.15 | 82.58 | 84.72 |
| Math | **92.00** | **91.42** | 69.47 | 70.14 | 68.74 | 68.06 | 79.52 | 80.52 | 74.31 | 77.19 | 91.01 | 90.17 |
| JacksonDatabind | **88.68** | **88.64** | 71.11 | 72.19 | 69.51 | 68.17 | 75.65 | 73.92 | 75.41 | 76.71 | 81.34 | 81.29 |
| JacksonCore | **92.73** | **92.48** | 70.04 | 68.10 | 70.46 | 71.50 | 80.61 | 81.94 | 70.98 | 71.53 | 88.19 | 87.92 |
| Total | **91.40** | **90.82** | 70.57 | 71.83 | 70.20 | 70.55 | 76.97 | 78.34 | 76.17 | 76.88 | 88.25 | 87.43 |

Table 6. Statistics for RQ1, comparing with Test2Vec classifier-TP

| | FFR | | APFD | |
|---|---|---|---|---|
| | P-Value | Effect-Size | P-Value | Effect-Size |
| Line Coverage | $1.22 \times 10^{-6}$ | 0.94 | 0.0019 | 0.87 |
| Branch Coverage | $8.80 \times 10^{-8}$ | 0.96 | 0.0019 | 0.87 |
| One-Hot | $1.48 \times 10^{-6}$ | 0.97 | 0.0019 | 0.87 |
| LSTM-based | $1.22 \times 10^{-6}$ | 0.94 | 0.0017 | 0.84 |
| CodeBERT | $3.98 \times 10^{-6}$ | 0.93 | 0.019 | 071 |

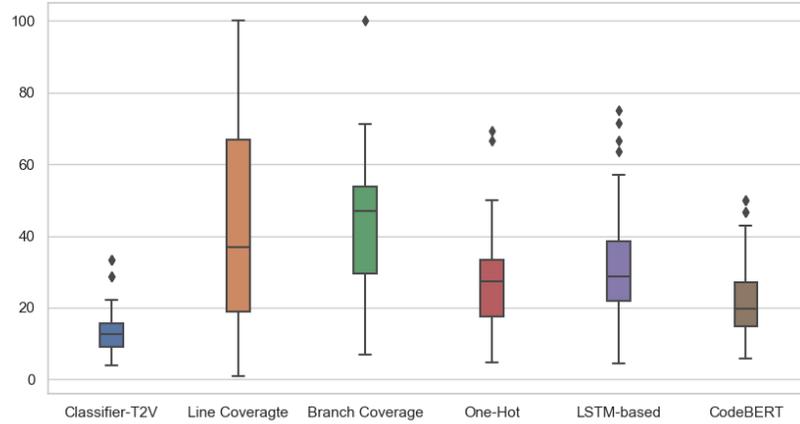

Fig. 7. Boxplots of the normalized ranks for the first failing test case for RQ1, prioritized by each TP technique.

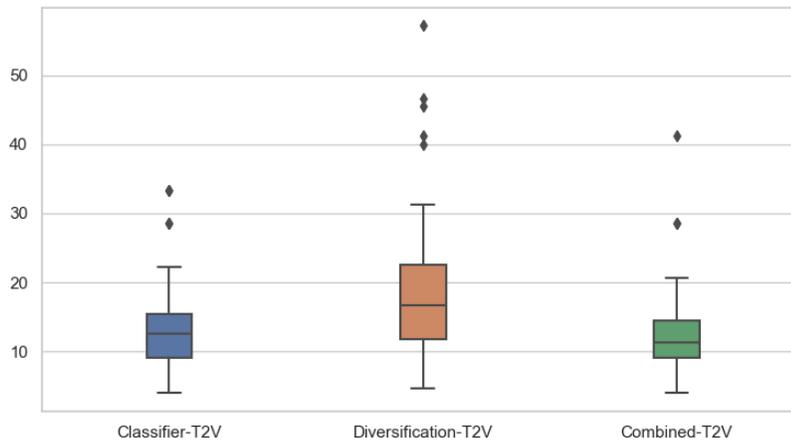

Fig. 8. Boxplots of the normalized ranks for the first failing test case for RQ2, prioritized by each TP technique.

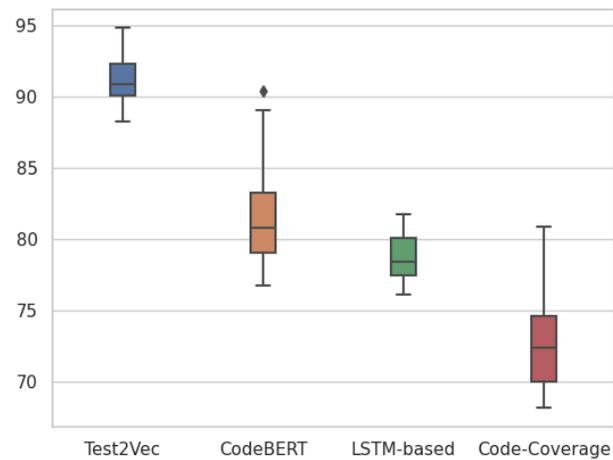

Fig. 9. Boxplots of the APFD, prioritized by each TP technique.

Tabel 5 also shows the APFD for all execution traces (including failing traces generated by mutation) among all 50 versions under study. the APFD for the Test2Vec classifier-TP has an average of 90.82 and a median of 91.40. In comparison, the line coverage(LC)-based TP has an average of 71.83 and a median of 70.57, and the branch coverage(BC)-based TP technique has an average of 70.55 and a median of 70.20. The relative improvement of the Test2Vec over the coverage-based TPs are: 26.42% (average) and 29.51% (median) over LC-based TP; and 28.72% (average) and 30.19% (median) over BC-based TP.

Figures 7, 8, and 9 report the whole distributions, as boxplots, of the normalized rank and the APFD, after repeating the experiment 30 times (by selecting randomly from test cases with a tied score) for all versions in every project, as boxplots. As the figure shows, Test2Vec classifier-TP not only offers better median and average test case ranking but also is more reliable, given the smaller variation over the randomized runs and versions.

Running the statistical tests (Wilcoxon Signed-Rank) on the APFD results shows that the effect size and p-value, when we are comparing the Test2Vec classifier-TP to the LC-TP, are 0.94 and $1.3 \times 10^{-6}$. When we Compare Test2Vec classifier-TP and the BC-TP results, the effect size is 0.96 and the p-value is $3.8 \times 10^{-7}$.

Table 3 also shows that, surprisingly, LC-based TP outperforms BC-based TP. This seems to contradict the fact that BC subsumes statement coverage. However, coverage subsumption does not necessarily translate into a better TP. This might simply be due to extra information (noise) in the BC that is harming the TP rather than being beneficial.

It also indicates that the representation itself is important; simply adding richer information (BC vs LC) does not guarantee improved performance.

**RQ1-2) Comparing with state-of-art code and execution trace embeddings:**

Table 3 also shows the FFR for the one-hot, CodeBERT, and LSTM-based TP methods. The one-hot TP method has an average of 27.27 and a median of 27.50, the CodeBERT TP technique has an average of 21.69 and a median of 19.09, and the LSTM-based method has an average of 31.41 and a median of 28.57. The relative improvement of the Test2Vec over the encoding and embedding baseline TPs are: 54.16% (average) and 53.24% (median) over One-Hot TP; 34.54% (average) and 40.7% (median) over CodeBERT TP; and 56.25% (average) and 59.06% (median) over LSTM-based TP.

For the APFD experiment, as Table 5 suggests, the APFD for the one-hot TP method has an average of 78.34 and a median of 76.97, the CodeBERT TP technique has an average of 87.48 and a median of 88.25, and the LSTM-based method has an average of 76.88 and a median of 76.17. This means that the relative improvement of the Test2Vec over the encoding and embedding baseline TPs are: 15.92% (average) and 18.75% (median) over One-Hot TP; 3.81% (average) and 3.57% (median) over CodeBERT TP; and 18.12% (average) and 20.00% (median) over LSTM-based TP.

Statistical tests results when we are comparing the Test2Vec classifier-TP with the One-Hot, CodeBERT, and LSTM-based one are effect size 0.97 and p-value $7.4 \times 10^{-7}$; effect size 0.93 and p-value $2.7 \times 10^{-7}$; and effect size 0.94 and p-value $4.4 \times 10^{-8}$, respectively.

The results show that (a) a simple encoding such as One-Hot cannot capture much relevant information from traces, for the TP task and that (b) CodeBert outperforms the LSTM-based embedding. This is interesting since the LSTM-based approach is specialized for execution traces and testing but it can not compete with off-the-shelf CodeBert. Furthermore, the results show that (c) our proposed embedding outperforms alternatives. This is notable, in particular since the advanced embedding models (CodeBERT and LSTM) use the same TP algorithms as Test2Vec (they all use the same pooling and softmax layers for calculating the ranks), and they are trained, validated, and tested on the same dataset splits. Thus the differences in the final results are due to the architecture of our proposed embedding.

Taken together, the RQ1 results show that the default Tes2Vec setup (Test2Vec classifier-TP) is significantly better than the coverage-based TP and state-of-art embedding TP techniques we compared it to. Comparing Test2vec classifier-TP with the best coverage-based (LC) and the best embedding (CodeBERT), the relative improvement regarding FFR (at the test suite level) are 68.89% (average) and 66.15% (median) over LC; and 40.71% (average) and 34.52% (median) over CodeBERT. Furthermore, in terms of APFD (in the whole project-version level), the relative improvements are 26.42% (average) and 29.51% (median) over LC; and 3.81% (average) and 3.57% (median) over CodeBERT.

Table 7. Statistics on the normalized ranks (FFRs) for RQ2, over all 42 versions under study (the high effect size indicates the left-side technique, per row, is outperforming the right-side technique).

|  | FFR | |
| --- | --- | --- |
|  | P-Value | Effect-Size |
| Classifier-TP vs Diversification-TP | 0.0027 | 0.80 |
| Combined-TP vs Diversification-TP | 0.00034 | 0.95 |
| Combined-TP vs Classifier-TP | 0.59 | 0.51 |

**RQ2) Which of the two prioritization heuristics (similarity to past failing tests or test diversification) is better to be used with Test2Vec for a test case prioritization task?**

To see if we can leverage the diversification for test case prioritization, we analyzed the results of the following sub-RQs:

**RQ2-1) Comparing diversification with classifier-TP:**

As discussed in section 5.4, diversification-TP is only evaluated using real faults (by reporting FFRs) and not mutants (i.e., no APFD). Table 4 shows the median (average) FFR results for both the classifier- and diversification-TPs.

The median FFRs for the Test2Vec diversification-TP method has an average of 19.03 and a median of 15.04. This means that the classifier-TP technique is dominating by a relative improvement of 32.42% (average) and 16.88% (median).

However, as shown in Figure 11, in some cases, diversification-TP can predict the failing test case in a better rank than classifier-TP. We can justify this with the fact that historical failures do not contain all the failing behaviors. Therefore, these failing behaviors may be detected as an anomaly within the latent space rather than by comparing them to historical failures. In other words, and for the specific projects we investigated, depending on the type of fault, about 76.2% of the time the similarity to past failures is a better heuristic, whereas about for 23.8% of the failures, detecting anomalous behavior within a test suite is a better heuristic. Thus, for other projects and circumstances, such as the length of the history, the strength/maturity of the test suite etc. one or the other of these heuristics might be preferable.

To give an example of how different faults can be better detected by one technique than the other, let's look at two sample failing tests from our dataset, illustrated in figure 10. The sub-figures show two statistics per sample: (a) how anomalous the failing test case is within its test suite (measured by its vector's distance to the centroid of the failing test suite – the more the better anomaly), and (b) how similar it is to the failing tests in the past (measured by its distance to the closest failing trace from historical versions – the less the closer to past failures). In figure 10.b, we see that for version 170 of the Chart project, using diversity to detect the failing test is a better heuristic, since the failing test's trace is not very similar to historical failing traces (the normalized distance to the closest failure is between 0.7 and 0.8,

Fig. 10. Illustrated distances of two sample test cases from the centroids of their test suites and their projects' historical failing tests, to motivate the Combined-TP approach. The four sub-figures show the distribution of the normalized distance of each trace from: their centroids in the latent space of their current test suite (a and c) and their nearest trace within the historical failing traces (b and d). The closer values to 1.0 in sub-figures a and c means the more anomalous behaviour. The closer value to 0.0 in sub-figures b and d means the closer behavior to past failing traces. Sub-figures (a) and (b) are from version 170 of the Chart project and sub-figures (c) and (d) are from the version 33 of Cli project.

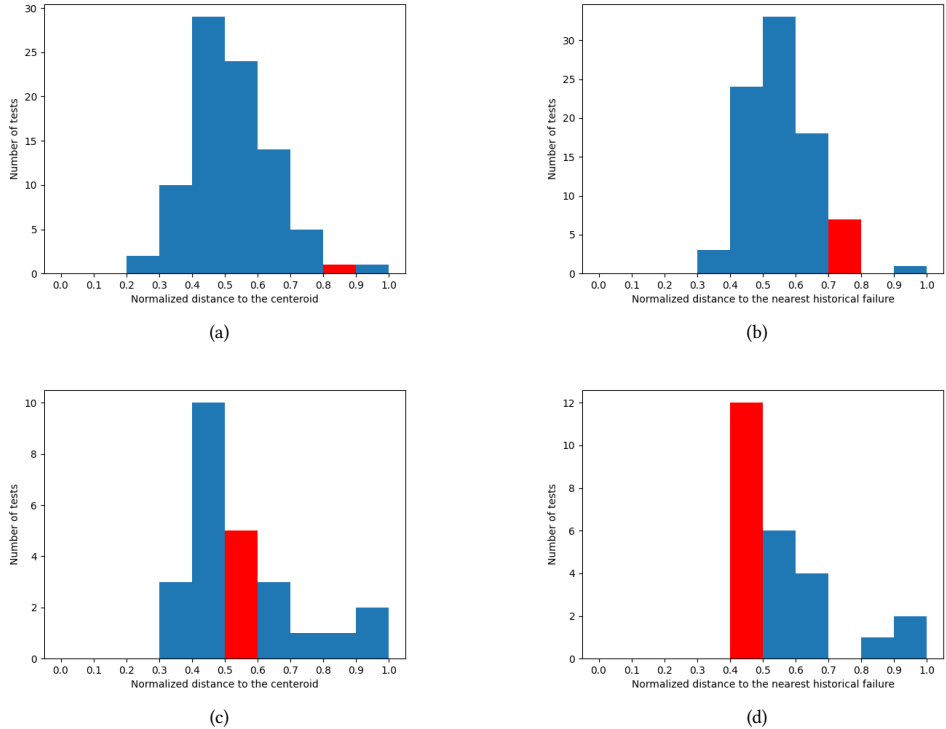

figure 10.b), but it looks quite like an anomaly in the current test suite (the normalized distance from the centroid is between 0.8 and 0.9, figure 10.a). On the other hand, a failing test case such as the one in version 33 of the Cli project, can be better detected by the classifier-TP, since the failing test's trace is quite close to at least one historical failure (the normalized distance to the closest failure is between 0.4 and 0.5, figure 10.d), but does not look like an anomaly within the test suite (the normalized distance from the centroid is between 0.5 and 0.6, figure 10.c)

Based on the above statistics and the provided example, given that the two types of failures are not necessarily overlapping, techniques that combines the two heuristics, such as the combined approach we evaluated in RQ2-2 might provide benefits.

**RQ2-2) Combining classifier- and diversification-TPs:**

As discussed, although the default Test2Vec classifier-TP outperforms the Test2Vec diversification-TP method, overall, there seem to be non-overlapping cases that the diversification heuristic is a better approach for detecting failing behaviors and thus test prioritization. In this sub-RQ, we compare the Test2Vec combined-TP with the default classifier-TP. As shown in Table 6, the average and median of FFR for the combined-TP method is 12.57 and 11.11. This means the combined-TP method slightly outperforms the classifier method by 2.25% (average) and 11.12% (median). However, Table 7 shows that the differences between the two techniques is not statistically significant.

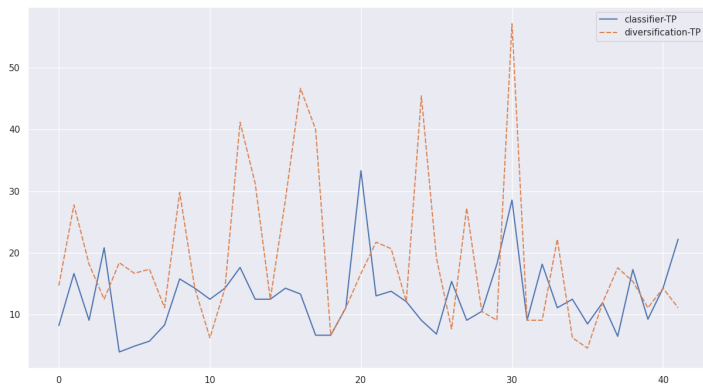

Fig. 11. Comparing the normalized rank of first failing test (FFR) for diversification- and classifier-TP.

The more interesting observation is perhaps the fact that Test2Vec combined-TP's median FFR, over 5 versions of each project, is better or the same as Test2Vec classifier-TP's results in 9 out of 10 projects. Basically there are three projects for which the combined-TPs' median FFRs are better, six projects for which the two techniques are equal, and only one case where combined-TP is not as effective. Looking into all 42 versions together, in 38 versions Test2Vec combined-TPs' FFR is better or equal to Test2Vec classifier-TP's FFR.

We also observe that the classifier used in the combined-TP method for selecting between the classifier- and diversification-TP can successfully detect the correct test case prioritization method in 83.33% of cases. Still, the simple selection method we have used can likely be improved upon; we leave it to future work to further investigate this.

Comparing our best proposal (Test2Vec combined-TP) with the best coverage-based and best alternative embedding from RQ1 also shows significant relative improvements. 59.25% and 41.80% in terms of FFR, over the best coverage and best alternative embedding, respectively.

Overall, although the improvements of Test2Vec combined-TP over Test2Vec classifier-TP is not significant, given that in 90.47% of the cases Test2Vec combined-TP is better or the same as the default version, we recommend it over the classifier-TP.

> Take together, RQ2 results show that the default Tes2Vec setup (Test2Vec classifier-TP) is significantly better than all alternatives including Test2Vec diversification-TP. However, the two heuristics for Test2Vec (i.e., similarity to failing tests from previous versions and being an anomaly in the current test suite) are complementary and the combined approach is slightly better than the default version.

### 5.7 Discussion on limitations of the proposed approach and the conducted study

The first limitation of our approach is that we need a training set of execution traces. Depending on the testing task and context, this may not be available. For instance, to use Test2Vec in test generation, we have to either start with an existing developer-written test suite or use an existing automated test generation tool to build one. Then Test2Vec can learn the representation of existing tests, before guiding the generation of new ones.

Another limitation is that, even with enough tests in the training set, if the number of real bugs are limited (which they usually are) we have to use mutation to create a (more) balanced dataset of fail/pass traces. But generating mutants and collecting traces for buggy tests in the mutated versions is time consuming. However, this is NOT a frequent task; in many cases it should be enough to do this once per project. As long as there aren't many changes to the project, the trained model can be reused. Therefore, we do not need to mutate it or collect traces, per change. The cost both of mutation and of creating embeddings can thus likely be amortized over many versions of a project, reducing the per-version cost. Future work is needed to investigate this and establish the cost versus performance trade-offs involved.

The training time for our approach is not negligible and can be substantial.Even if, as discussed above, this can be amortized over several versions, it is a limitation that over time, and multiple versions, the model's accuracy may decline.

Generally speaking, balancing the trade-off between the cost of re-training and still having an accurate model without re-training is a problem we did not address in this paper. In our experiments, we looked at the results with one time training (on 20 versions) and five times reuse. We did not observe a significant degradation over those five uses. But a more comprehensive study could investigate when and how to re-train the models. It also worth mentioning that retraining may affect both the accuracy of the embeddings as well as the downstream task procedure. In our experiments, the effect of retraining can be bolder on the classifier-TP compared to diversification-TP, given that classifier-TP's accuracy partly depends on how rich and relevant the historical failures still are. However, Diversification-TP is only affected in the embedding generation phase and the ranking procedure only depends on the current test suite.

One limitation of the experiment (not the approach itself) is we implemented the TPs using a greedy approach (same across the proposed approaches and baselines) for optimizing the rank list. More recent TP papers try to use optimization algorithms such as evolutionary algorithms [51]. Given that our focus was on the test representation and not the optimization approach, we did not explore this dimension. In other words, we argue that the representation and the optimization based on it are orthogonal choices and this paper only focuses on the former. Future work should investigate if optimization on top of the proposed embeddings can provide additional benefits. It would be beneficial to explore further many of the design choices in this study. For instance, one study can be designed to compare the effect of different diversity functions on Test2Vec diversification-TP results. Another study can be on exploring the idea of input parameter abstraction in more detail to come up with different alternatives to compare.

Another limitation of the experiment was that we applied our approach at the unit testing level by eliminating nested calls. By focusing on inter-class interaction, future works can apply this same approach to integration testing as well.

Finally, we used the CodeBERT for token embedding generation for outputs, method calls, and input sequences. This means that we have a 512 tokens limitation for each sequence. There are many works on the domain of NLP for handling larger sequuences [10, 67]. The same ideas can be applied on our problem as future works.

### 5.8 Validity threats

Regarding construct validity, in projects like Lang and Time, the traces' length varies from a few calls to thousands of calls, therefore, in the pre-processing step, we perform padding or truncation to get a fixed-length set of traces. This may cause information loss in very long traces, i.e., our representation of those test cases does not accurately measure what it claims (its behavior). However, these cases are rare and likely would not change the overall findings.

In terms of internal validity, we chose our evaluation metrics quite carefully! We wanted metrics that work on both real-world bugs as well as mutations. We also wanted to be able to evaluate our TPs both on the whole project version as well as test suite level. The the rank of first failing test is suitable for real-world faults, where we have only one fault in the class under test with very few failing test cases in the failing suite. But we had to normalize it since without normalization large and small test sets results where not comparable. The APFD, on the other hand, is used for the TP task when we use mutation analysis on the whole version's test cases, where we have multiple mutants (bugs) per class, resulting in multiple failing tests. Although the caveat of APFD was that we could not use APFD in RQ2 (since diversification in the presence of multiple mutants would not give any meaningful results; so many anomalies won't make sense), but at least for RQ1 we managed to explore the results from all aspects (real-world vs mutants AND test suite vs whole version test cases).

In terms of conclusion validity, we repeat our training on 10 different projects and report the statistical tests and effect size across multiple versions. We also repeat the TPs for 30 runs, whenever the TP had randomness in the process, per version. We then carefully reported both median and mean, as well as paired non-parametric statistical tests and effect sizes, to better analyze the differences and their statistical an practical significance.

Finally, in terms of external validity, although we tried to mitigate the threat by selecting multiple (10) projects from a benckmark dataset, our findings are limited to TP in the context of unit testing of Defect4J projects. So our findings can not be generalized for other types of test cases or other projects specially industrial ones, without further replications.

## 6 RELATED WORK

In this section, we discuss about some of the related works in representation learning in software engineering in general and in software testing in particular. Note that for the sake of brevity we don't not discuss all test case prioritization literature and only include them if they use a novel test representation. For more information on TP techniques, the readers can consult with recent surveys and reviews in this domain [39, 51, 66].

### 6.1 Sequence Learning and Embedding in Software Engineering

In software engineering, sequence learning and embedding has been used for tasks such as code completion[71], program repair [81], API learning [34] and code search [33]. The focus of learning in these works is on the source code [3] and their crucial drawback, in our context, is missing the execution information, which is one the most important concerns in testing.

There are several applications of sequence learning in software engineering in recent years [4, 16, 35, 61]. For example, Harer *et al.* [36] use Word2Vec on source code to predict possible bugs or security vulnerabilities of the program. While *et al.* [55] generate embedding using an CodeBERT model for program repair. Azcona *et al.* [8] use a simple vector representation model to characterize code, and Chen and Monperrus [17] use Word2Vec and Doc2Vec for program repair. In the category of customized embeddings for source code (neural program embedding), Code2Vec [6], Code2Seq [5], Flow2Vec [73], and CodeBERT [28] are the state-of-the-art. Though these model's underlying architectures defer, all are working on the program source code and miss the execution information of traces. Since CodeBERT is a pre-trained model (trained on a very large code base) and has outperformed the others, we selected it as our baseline embedding techniques.

In the context of representing dynamic executions, Henkel *et al.* [43] use semantic and syntactic abstracted traces from symbolic execution for C projects to train a Word2Vec embedding model. Their method, however, does not include input values in the training data due to the abstraction. Wang *et al.* [79] propose a dynamic neural program embedding for program repair. They show that using traces including the variables has better results than using the variables or the states traces. In a more recent work [80], for the task of learning representation vectors for the method's body, the authors show that a mix of symbolic and concrete execution traces outperform other existing approaches that ignore dynamic behavior of the program. These works are among our motivating works that though applied on different tasks but showed that dynamic traces including input data may be beneficial in testing.

## 6.2 Representation Learning for Software Testing

**Using coverage information:** Early works in the test case prioritization mostly used coverage data of test cases to represent each test case. Elbaum *et al.* [21] used multiple coverage-based representations (vary in granularity and considering additional or total coverage), in the TP task. They conclude that techniques with finer granules typically outperform others. D. Nardo *et al.* [20] also showed that coverage-based techniques that use additional coverage with more detailed coverage criteria are more effective at detecting faults.

**Using text similarity:** In this category, each test case is considered as a string by looking at its source code, and the similarity or distance of test cases is measured by applying simple string distance metrics (without applying any encoding or embedding). For instance, Miranda **et al.** [58] used string distance metrics, such as Hamming distance and Levenshtein diversity, to measure similarity/distance between test cases.

**Using static data of test cases:** This category contains works that use static data related to test cases such as source code, specification models, comments, and ASTs in order to represent test cases. For example, Hemmati *et al.* [40] encode tests based on specification coverage of each test case on the manually drafted state machine representation of the system. Mondal *et al.* [59] used one-hot encoding to represent each test case based on a set of method calls that are extracted from each test case script, without executing the code. Thomas *et al.*[74] uses a topic modeling approach on the source code and comments.

**Using input data:** These works usually try to represent test cases based on one or a sequence of inputs. In other words, they are black-box fuzzing approaches that only consider test inputs. For example, Godefroid *et al.* [30] have leveraged neural statistical models to automatically infer input grammars for Fuzz testing.

**Using dynamic traces of test cases:** Another approach for representing test cases is using dynamic data, such as execution trace. Using dynamic data for software testing is not explored enough by researchers. Noor and Hemmati [65] used dynamic execution trace of the program to represent a test case by applying a one-hot encoding that only consider method calls. MK Ramanathan *et al.* [70] represents each test case with the sequence of memory operations and their values at the run time execution of the test case. In order to find the similarity of tests within a test suite, these sequences are then used to create a dissimilarity graph. Tsimpourlas *et al.* [77] used recurrent NN and execution trace of the programs for classifying test cases as pass or fail. We used this approach as one of our baselines in this paper.

# 7 CONCLUSION AND FUTURE WORK

In this paper, we proposed a novel neural embedding model called Test2Vec, that embeds test cases as numerical vectors which can, in turn, improve performance on downstream testing tasks such as test prioritization. The flexibility of neural embeddings enables representation of rich test case information such as the ordering of method calls as well as specific test input and output values.

Empirical evaluation, on ten projects from Defects4J, show that our approach outperforms traditional code coverage metrics, classic diversity-based encoding, state-of-the-art neural program embedding models, and an embedding with other sequence learning approaches, on a test prioritization task.

Future work should investigate the effect of including even more, relevant information in the execution traces as well as study the use of Test2Vec embeddings for other downstream testing tasks such as automated test generation. We will also investigate the applicability of this approach, by using other dynamic representations of the program behaviour, on system, security, and performance testing.